\documentclass[twocolumn,showpacs,preprintnumbers,superscriptaddress,amsthm,amsmath,amssymb]{revtex4}
\usepackage{graphicx}
\usepackage{dcolumn}
\usepackage{bm}
\usepackage{color}
\usepackage{booktabs, longtable}
\usepackage{multirow}
\usepackage{amsmath}
\usepackage{CJK}
\usepackage{ltxtable}
\usepackage{rotating}
\usepackage{dcolumn,siunitx,mathptmx,booktabs}
\usepackage[colorlinks,linkcolor=blue,anchorcolor=blue,citecolor=red]{hyperref}

\begin{document}

\title{Green's function method for the spin and pseudospin symmetries \\ in the single-particle resonant states}

\author{Ting-Ting Sun}
\email{ttsunphy@zzu.edu.cn}
\affiliation{School of Physics and Engineering, Zhengzhou University, Zhengzhou 450001, China}

\author{Wan-Li Lu}
\affiliation{School of Physics and Engineering, Zhengzhou University, Zhengzhou 450001, China}

\author{Long Qian}
\affiliation{School of Physics and Engineering, Zhengzhou University, Zhengzhou 450001, China}

\author{Yu-Xiao Li}
\affiliation{School of Physics and Engineering, Zhengzhou University, Zhengzhou 450001, China}
\date{\today}

\begin{abstract}
We investigate the spin and pseudospin symmetry in the single-particle resonant states by solving the Dirac equation containing a Woods-Saxon potential with Green's function method. Taking double-magic nucleus $^{208}$Pb as an example, three spin doublets $3d$, $2h$, and $1j$ and three pseudospin doublets $3\tilde{p}$, $1\tilde{i}$, and $1\tilde{j}$ are obtained for the single-neutron resonant states. By analyzing the energy splittings, we find that the threshold effect plays an important role in resonant pseudospin doubles.
Besides, there is a reversed level structure of pseudospin doublets in the continuum.
Differently, all the width splittings of either the spin doublets or the pseudospin doublets are systematically positive and the splittings are very small except $1\tilde{j}$ doublet. Further studies show that the splittings of the energies and widths for the resonant (pseudo)spin doublets are independent. Besides, the similarity properties of the wave functions of the spin and pseudospin doublets still maintain well in resonant states.
\end{abstract}

\pacs{25.70.Ef, 21.10.Pc, 21.60.Jz, 21.10.Gv}
\maketitle

\section{Introduction}
\label{sec:intr}

Symmetries in the single-particle spectrum of atomic nuclei are of great importance on nuclear structure and have been extensively discussed in the literature~(see Refs.~\cite{PhysRep2005Ginocchio_414_165,PhysRep2015HZLiang_570_1}~and references therein). The breaking of spin symmetry (SS), i.e., the remarkable spin-orbit (SO) splitting for the spin doublets ($n,l,j=l \pm 1/2$) caused by the SO potential, lays the foundation to explain the traditional magic numbers in nuclear physics~\cite{PhysRev1949Haxel_75_1766,PhysRev1949Mayer_75_1969}. The conservation of pseudospin symmetry (PSS), i.e., the quasi-degeneracy between two single-particle states with quantum numbers ($n,l,j=l+1/2$) and ($n-1,l+2,j=l+3/2$) redefined by the pseudospin doublets ($\tilde{n}=n$, $\tilde{l}=l+1$, $j=\tilde{l}\pm 1/2$)~\cite{NPA1969Hecht_137_129,PLB1969Arima_30_517}, has been used to explain a number of phenomena in nuclear structures, including deformation~\cite{PhysScr1982Bohr_26_267}, superdeformation~\cite{PRL1987Dudek_59_1405}, magnetic moment~\cite{PRL1990Nazarewicz_64_1654}, identical rotational bands~\cite{PRL1990Byrski_64_1650} and so on.

Since the recognition of PSS in the nuclear spectrum, comprehensive efforts have been made to understand its origin. In 1997, Ginocchio made a substantial progress and clearly pointed out that PSS is a relativistic symmetry in the Dirac Hamiltonian and becomes exact when the scalar and vector potentials satisfying $\Sigma(r) \equiv S(r)+V(r)=0$~\cite{PRL1997Ginocchio_78_436}. He also revealed that the pseudo-orbital angular momentum $\tilde{l}$ is nothing but the orbital angular momentum of the lower component of the Dirac wave function~\cite{PRL1997Ginocchio_78_436} and
there are certain similarities in the relativistic single-nucleon wave functions of the corresponding pseudospin doublets~\cite{PRC1998Ginocchio_57_1167}. With a more general condition, Meng \textit{et al}. pointed out that $d \Sigma(r)/d r=0$ can be approximately satisfied in exotic nuclei with highly diffuse potentials~\cite{PRC1998JMeng_58_628,PRC1999JMeng_59_154} and the onset of the pseudospin
symmetry to a competition between the pseudo-centrifugal barrier (PCB) and the pseudospin-orbit (PSO) potential. Afterwards, the SS and PSS in nuclear spectra have been studied extensively such as PSS in deformed nuclei~\cite{PRC1998Lalazissis_58_R45,PRC1999Sugawara_58_R3065,PRC2000Sugawara_62_054307,PRC2002Sugawara_65_054313,
PRC2004Ginocchio_69_034303,EPJA2012YWSun_48_18}, PSS and SS in hypernuclei~\cite{CPL2009CYSong_26_122102,CPC2010CYSong_34_1425,CPL2011CYSong_28_092101,PRC2017Sun_96_044312,JPG2017Lu_44_125104,NPR2019Lu},
SS in anti-nucleon spectra~\cite{PRL2003SGZhou_91_262501,PRC2005Mishustin_71_035201,EPJA2006He_28_265,EPJA2010HZLiang_44_119,
PRC2010Lisboa_81_064324}, perturbative interpretation of SS and PSS~\cite{PRC2002Alberto_65_034307,IJMPD2004Alberto_13_1447,PRC2010Lisboa_81_064324,PRC2011Liang_83_041301,CPC2011FQLi_35_825}, and PSS in supersymmetric quantum mechanics~\cite{PRL2004Leviatan_92_202501,PRL2009Leviatan_103_042502,PRC2013Liang_87_014334,PRC2013Shen_88_024311}.

In recent years, the study of single-particle resonant states has attracted great attentions due to the essential roles in the exotic nuclei with unusual $N/Z$ ratios. In exotic nuclei, the neutron or the proton Fermi surface is very close to the continuum threshold and the valence nucleons can be easily scattered to the single-particle resonant states in the continuum due to the pairing correlations and the couplings between the bound states and the continuum become very important~\cite{PRL1998MengJ_80_460,PRC1996Dobaczewski_53_2809,PRL1996Meng_77_3963,PRL1997Poschl_79_3841,
PPNP2006JMeng_57_470}. For example, the self-consistent relativistic continuum Hartree-Bogoliubov (RCHB) calculations suggested that the neutron halo in the first observed halo nucleus $^{11}$Li is formed by scattering Cooper pairs to the $2s_{1/2}$ orbit in the continuum~\cite{PRL1996Meng_77_3963}. Similarly, the predicted giant halos in exotic Zr and Ca isotopes by RCHB theory are formed with more than two valence neutrons scattered as Cooper pairs to the continuum~\cite{PRL1998MengJ_80_460,PRC2002MengJ_65_041302,SCP2003ZhangSQ_46_632}.
By including only the contribution of the resonant states, the giant halos were also reproduced by the relativistic mean-field calculations with pairing treated by the BCS method~\cite{PRC2003Sandulescu_68_054323}.
Moreover, taking the deformed relativistic Hartree Bogoliubov (DRHB) theory in continuum, the studies of deformed halo nuclei show that the existence and the deformation of a possible neutron halo depend essentially on the quantum numbers of the main components of the single-particle orbitals in the vicinity of the Fermi surface~\cite{PRC2010Zhou_82_011301,PRC2012Li_85_024312,PRC2012Chen_85_067301,CPL2012Li,PLB2018Sun}. Therefore, the properties of the resonant states close to the continuum threshold are essential for the investigation of exotic nuclei.

Thus, the study of the PSS and SS symmetry in the single-particle resonant states should also be of great interests. Until now, there are already some investigations of the PSS in the single-particle resonant states. As early as around 2000, the PSS and SS in nucleon-nucleus and nucleon-nucleon scattering have been investigated~\cite{PRL1999Ginocchio_82_4599,PRC2000HLeeb_62_024602,PRC2002Ginocchio_65_054002,PRC2004HLeeb_69_054608}. In 2004, Zhang \textit{et~al.} confirmed that the lower components of the Dirac wave functions for the resonant pseudospin doublets also have similarity properties~\cite{CPL2004Zhang_21_632}. In 2005 and 2006, Guo \textit{et~al.} investigated the dependence of the pseudospin breaking for the resonant states on the shape of the mean-field potential in a  Woods-Saxon form~\cite{PRC2005JYGuo_72_054319} as well as on the ratio of neutron and proton numbers~\cite{PRC2006JYGuo_74_024320}. In 2012, Lu \textit{et~al.} gave a rigorous justification of PSS in single-particle resonant states and shown that the PSS in single-particle resonant states is also exactly conserved under the same condition for the PSS in bound states, i.e., $\Sigma(r)=0$ or $d\Sigma(r)/dr=0$~\cite{PRL2012BNLu_109_072501}.

During the past years, many approaches have been developed or adopted to study the single-particle resonant states, such as $R$-matrix theory~\cite{PR1947Wigner_72_29,PRL1987Hale_59_763}, $K$-matrix theory~\cite{PRC1991Humblet_44_2530}, $S$-matrix theory~\cite{Book1972Taylor-ScatteringTheor,PRC2002CaoLG_66_024311}, Jost function approach~\cite{PRL2012LuBN_109_072501, PRC2013LuBN_88_024323}, and the scattering phase shift (SPS)  method~\cite{Book1972Taylor-ScatteringTheor,PRC2010LiZP_81_034311,SCP2010Li_53_773}, which are based on the conventional scattering theories. Meanwhile, some techniques for bound states have been extended for the single-particle resonant states, such as the analytical continuation of the coupling constant~(ACCC) method~\cite{Book1989Kukulin-TheorResonance,PRC1997Tanaka_56_562,PRC1999Tanaka_59_1391,PRC2000Cattapan_61_067301,CPL2001YangSC_18_196,PRC2004ZhangSS_70_034308,PRC2005Guo_72_054319,EPJA2007SSZhang_32_43,PRC2012SSZhang_86_032802,EPJA2012SSZhang_48_40,EPJA2013SSZhang_49_77,PLB2014SSZhang_730_30,PRC2015Xu_92_024324}, the real stabilization method~(RSM)~\cite{PRA1970Hazi_1_1109,PRL1993Mandelshtam_70_1932,PRA1999Kruppa_59_3556,NPA2004Hagino_735_55,PRC2008ZhangL_77_014312}, the complex scaling method~(CSM)~\cite{PR1983Ho_99_1,PRC1986Gyarmati_34_95,PRC1988Kruppa_37_383,PRL1997Kruppa_79_2217,PRC2006Arai_74_064311,CPC2010Guo_181_550,PRC2010JYGuo_82_034318,PRC2014ZLZhou_89_034307,PRC2015MShi_92_054313,PRC2012QLiu_86_054312,PRC2014Shi_90_034319}, the complex-scaled Green's function (CGF) method~\cite{PLB1998Kruppa_431_237,PTP2005Suzuki_113_1273,EPJA2017Shi_53_40}, and the complex momentum representation~(CMR) method~\cite{NPA1968Berggren_109_265,PRC1978Kwon_18_932,JPA1979Sukumar_12_1715,PRC2006Hagen_73_034321,PRL2016Li_117_062502,PRC2017Fang_95_024311,PRC2017Tian_95_064329}.
Combined with the relativistic mean-field (RMF) theory, which has achieved great successes in describing many systems, such as the stable and exotic nuclei~\cite{PPNP2006JMeng_57_470,ANP1986Serot_16_1,RPP1989Reinhard_52_439,PPNP1996Ring_37_193,PR2005Vretenar_409_101}, hypernuclei~\cite{PRC2016Sun,PRC2018Liu,NPR2019Liu}, neutron stars~\cite{CPC2018Sun,PRD2018Sun,PRD2019Sun}, and $r$-process simulations~\cite{Sun2008PRC, Niu2009PRC, Xu2013PRC, Niu2013PLB, Zheng2014PRC}, some of the above methods for single-particle resonant states have been introduced to investigate the resonances. As examples, the RMF-ACCC approach is used to give the energies and widths~\cite{CPL2001YangSC_18_196} as well as the wave functions~\cite{PRC2004ZhangSS_70_034308,EPJA2007SSZhang_32_43} of resonant states. Similar applications for the Dirac equations with square well, harmonic oscillator, and Woods-Saxon potentials can be seen in Ref.~\cite{CPL2004Zhang_21_632}. The RMF-RSM approach is introduced to study the single-particle resonant states in $^{120}$Sn~\cite{PRC2008ZhangL_77_014312}. The RMF-CSM is developed to describe the single-particle resonant states in spherical~\cite{PRC2010JYGuo_82_034318,PRC2014ZLZhou_89_034307} and deformed nuclei~\cite{PRC2012QLiu_86_054312,PRC2014Shi_90_034319}. The single-particle resonant states in deformed nuclei have been investigated by the coupled-channel approach based on the scattering phase-shift method as well~\cite{PRC2010LiZP_81_034311}.

Green's function (GF) method~\cite{PRB1992Tamura_45_3271,PRA2004Foulis_70_022706,Book2006Eleftherios-GF} is also an efficient tool for studying the single-particle resonant states with the following advantages: treating the discrete bound states and the continuum on the same footing; providing both the energies and widths for the resonant states directly; and having the correct asymptotic behaviors for the wave functions. Nonrelativistically and relativistically, there are already many applications of the GF method in the nuclear physics to study the contribution of continuum to the ground states and excited states. Nonrelativistically, in the spherical case, in 1987, Belyaev \textit{et~al.} constructed the Green's function in the Hartree-Fock-Bogoliubov (HFB) theory in the coordinate representation~\cite{SJNP1987Belyaev_45_783}. Afterwards, Matsuo applied this Green's function to the quasiparticle random-phase approximation (QRPA)~\cite{NPA2001Matsuo_696_371}, which was further used to describe the collective excitations coupled to the continuum~\cite{PTPS2002Matsuo_146_110, PRC2005Matsuo_71_064326,NPA2007Matsuo_788_307,PTP2009Serizawa_121_97,PRC2009Mizuyama_79_024313,PRC2010Matsuo_82_024318,PRC2011Shimoyama_84_044317}, microscopic structures of monopole pair vibrational modes and associated two-neutron transfer amplitudes in neutron-rich Sn isotopes~\cite{PRC2013Shimoyama_88_054308}, and neutron capture reactions in the neutron-rich nuclei~\cite{PRC2015Matsuo_91_034604}. Recently, Zhang \emph{et~al.} developed the fully self-consistent continuum Skyrme-HFB theory with GF method~\cite{PRC2011ZhangY_83_054301,PRC2012YZhang_86_054318}, which was further extended for odd-A nuclei~\cite{arXivSun2013}. In the deformed case, in 2009, Oba \emph{et~al.} extended the continuum HFB theory to include deformation on the basis of a coupled-channel representation and explored the properties of the continuum and pairing correlation in deformed nuclei near the neutron drip line~\cite{PRC2009Oba_80_024301}. Relativistically, in the spherical case, in Refs.~\cite{PRC2009Daoutidis_80_024309,PRC2010DYang_82_054305}, the fully self-consistent relativistic continuum random-phase-approximation (RCRPA) was developed with the Green's function of the Dirac equation and used to study the contribution of the continuum to nuclear collective excitations. Recently, we have developed the continuum covariant density functional theory (CDFT) based on the GF method and calculated the accurate energies and widths of the single-neutron resonant states for the first time~\cite{PRC2014TTSun_90_054321}. This method has been further extended to describe single-particle resonances for protons~\cite{JPG2016TTSun_43_045107} and $\Lambda$ hyperons~\cite{PRC2017Ren_95_054318}. In 2016, further containing pairing correlation, we developed Green's function relativistic continuum Hartree-Bogoliubov (GF-RCHB) theory, by which the continuum was treated exactly and the giant halo phenomena in neutron-rich Zr isotopes were studied~\cite{Sci2016Sun_46_12006}.

In this work, Green's function method is applied to investigate the (pseudo)spin symmetry in the single-particle resonance states. The paper is organized as following. We give the formulations of the Green's function method in Sec.~\ref{sec:form}. After the numerical details in Sec.~\ref{sec:Numer}, we present results and discussions in Sec.~\ref{sec:resu}. Finally, a brief summary is given in Sec.~\ref{sec:Sum}.

\section{Formalism}
\label{sec:form}

In the RMF theory~\cite{ANP1986Serot_16_1, RPP1989Reinhard_52_439, PPNP1996PRing_37_193,PR2005Vretenar_409_101, PPNP2006MengJ_57_470}, nucleons are described as Dirac spinors moving in a mean-field potential, and the corresponding Dirac equation is
\begin{equation}
  [{\bm \alpha}\cdot{\bm p}+V({\bm r})+\beta(M+S({\bm r}))]\psi({\bm r})=\varepsilon \psi({\bm r}),
\label{Eq:DiracEq}
\end{equation}
where ${\bm \alpha}$ and $\beta$ are the Dirac matrices, $M$ is the nucleon mass, and $S({\bm r})$ and $V({\bm r})$ are the scalar and vector potentials, respectively.

In the spherical case, the Dirac spinors of nucleons can be expanded as
\begin{equation}
    \psi({\bm r}) =\frac{1}{r}
    \left(\begin{array}{c}
    i G_{n\kappa}(r) Y_{jm}^l(\theta,\phi) \\
    - F_{\tilde{n}\kappa}(r) Y_{jm}^{\tilde{l}}(\theta,\phi) \\
    \end{array}\right)  ,
    \label{EQ:RWF}
\end{equation}
where $G_{n\kappa}(r)/r$ and $F_{\tilde{n}\kappa}(r)/r$ are the upper and lower components of the radial wave functions with $n$ and $\tilde{n}$ the numbers of radial nodes, $Y_{jm}^l(\theta,\phi)$ is the spinor spherical harmonic, $\tilde{l}=l-\mathrm{sign}(\kappa)$ is pseudo-orbital angular momentum, and the quantum number $\kappa$ is defined as $\kappa=(-1)^{j+l+1/2}(j+1/2)$.

With the radial wave functions, the Dirac equation (\ref{Eq:DiracEq}) is reduced as
\begin{equation}
   \left(\begin{array}{cc}
    \Sigma & -\displaystyle{\frac{d}{dr} + \frac{\kappa}{r}}  \\
    \displaystyle{\frac{d}{dr}+\frac{\kappa}{r}}  & \Delta-2M    \\
   \end{array}\right)
   \left(\begin{array}{c}
   G \\
   F \\
   \end{array}\right)
   = \varepsilon
   \left(\begin{array}{c}
   G \\
   F \\
   \end{array}\right),
   \label{EQ:RDirac}
\end{equation}
where the mean-field potentials $\Sigma = V + S$ and $\Delta = V - S$. Note that the single-particle energy $\varepsilon$ in Eq.~(\ref{EQ:RDirac}) is shifted with respect to $M$ compared to that in Eq.~(\ref{Eq:DiracEq}).

In order to study the SS and PSS of single-particle resonant states with the RMF theory, the Dirac equation governing the motion of nucleons will be examined by the Green's function method, in which the relativistic single-particle Green's function should be constructed in coordinate space according to the definition,
\begin{equation}
[\varepsilon-\hat{h}(\bm r)]\mathcal{G}(\bm r,\bm r';\varepsilon)=\delta(\bm r-\bm r'),%
\end{equation}
where $\hat{h}(\bm r)$ is the Dirac Hamiltonian.
Taking the Green's function, the single-particle spectra including the bound and resonant states can be treated on the same footing by the density of states $n(\varepsilon)$~\cite{PRC2014TTSun_90_054321},
\begin{equation}
  n(\varepsilon)=\sum_{i}\delta(\varepsilon-\varepsilon_{i}),
  \label{EQ:DOS}
\end{equation}
where $\varepsilon_{i}$ is the eigenvalue of the Dirac equation, $\sum_{i}$ includes the summation for the discrete states and the integral for the continuum, and $n(\varepsilon)d\varepsilon$ gives the number of states in the interval $[\varepsilon, \varepsilon+d\varepsilon]$. For the bound states, the density of states $n(\varepsilon)$ exhibits discrete $\delta$-function at $\varepsilon=\varepsilon_{i}$, while in the continuum $n(\varepsilon)$ has a continuous distribution.

With a complete set of eigenstates~$\psi_{i}(\bm r)$~and eigenvalues~$\varepsilon_i$, the Green's function for nuclei can be represented as
\begin{equation}
  \mathcal{G}(\bm r,\bm r';\varepsilon)= \sum_i\frac{\psi_{i}(\bm r)\psi_{i}^\dagger(\bm r')}{\varepsilon-\varepsilon_i},
  \label{EQ:GF}
\end{equation}
where~$\Sigma_i$~is summation for the discrete states and integral for the continuum explicitly. Green's function in Eq.~(\ref{EQ:GF}) is analytic
on the complex energy plane with the poles at eigenvalues~$\varepsilon_i$. Corresponding to the upper $G_{i}({\bm r})$ and lower $F_{i}({\bm r})$ components of the Dirac spinor~$\psi_{i}({\bm r})$, the Green's function for the Dirac equation is in a form of a~$2\times2$~matrix,
\begin{equation}
 \mathcal{G}(\bm r,\bm r';\varepsilon)=
 \left(
        \begin{array}{cc}
          \mathcal{G}^{(11)}(\bm r,\bm r';\varepsilon)&  \mathcal{G}^{(12)}(\bm r,\bm r';\varepsilon) \\
          \mathcal{G}^{(21)}(\bm r,\bm r';\varepsilon) &  \mathcal{G}^{(22)}(\bm r,\bm r';\varepsilon)\\
        \end{array}
 \right).
\end{equation}

In the spherical case, Eq.~(\ref{EQ:DOS}) becomes
\begin{equation}
  n(\varepsilon)=\sum_{\kappa}n_{\kappa}(\varepsilon),
  \label{EQ:RDOS}
\end{equation}
where $n_{\kappa}(\varepsilon)$ is the density of states for a block characterized by the quantum number $\kappa$.
By introducing an infinitesimal imaginary part $``i\epsilon"$ to energy $\varepsilon$, it can be proved that the density of states can be obtained by integrating the imaginary part of the Green's function over the coordinate space, and in the spherical case, it is~\cite{PRC2014TTSun_90_054321}
\begin{eqnarray}
  n_{\kappa}(\varepsilon)&=&-\frac{2j+1}{\pi }\int d{r}{\rm Im}[\mathcal{G}_{\kappa}^{(11)}({r},{r};\varepsilon+i\epsilon)\nonumber \\
  ~&&+\mathcal{G}_{\kappa}^{(22)}({r},{ r};\varepsilon+i\epsilon)].
\label{EQ:DOS-lj}
\end{eqnarray}
Moreover, with this infinitesimal imaginary part $``i\epsilon"$, the density of states for discrete single-particle states in shape of $\delta$-function~(no width) is simulated by a Lorentzian function with the full-width at half-maximum (FWHM) of $2\epsilon$.

For the construction of the Green's function of Dirac equation and detailed formalism in GF-RMF model, the readers are referred to Ref.~\cite{PRC2014TTSun_90_054321}.

\begin{figure*}[htbp]
  \includegraphics[width=0.9\linewidth]{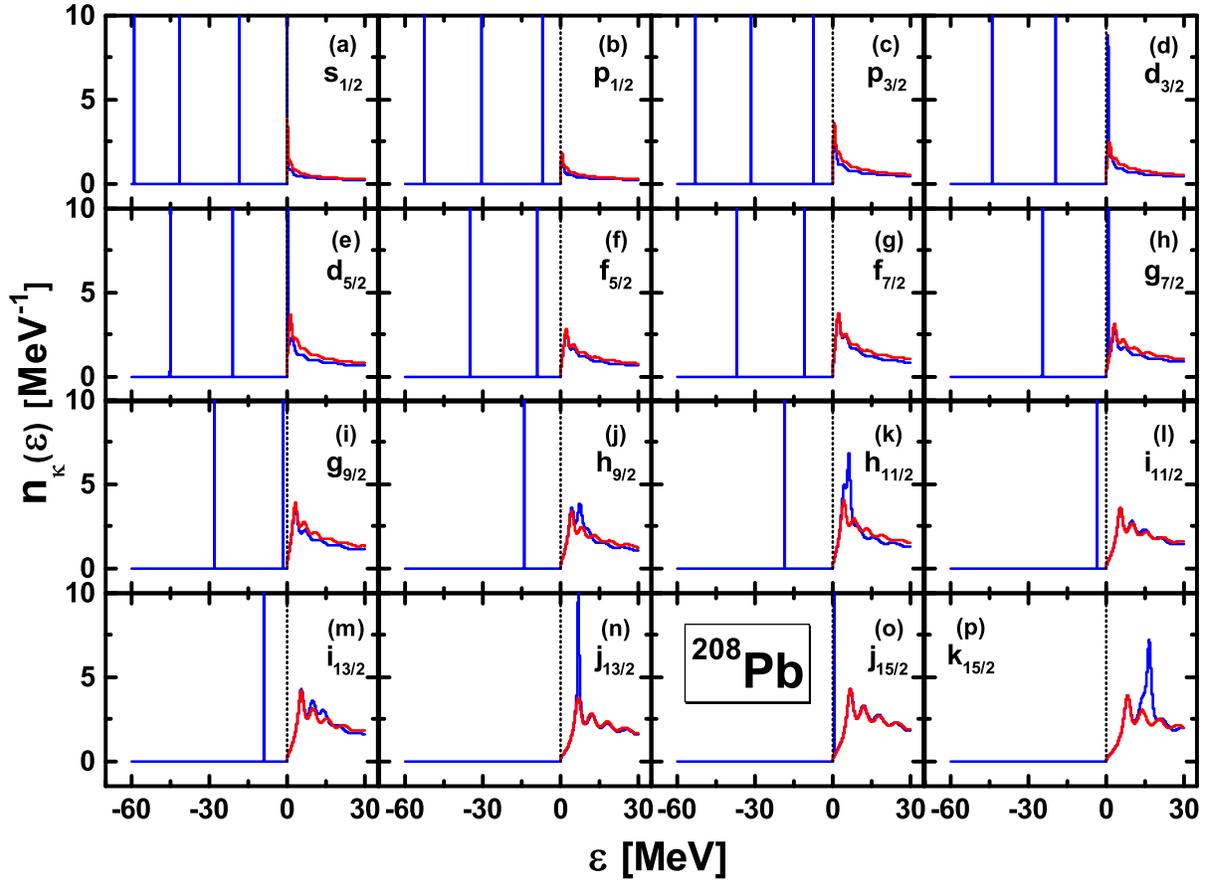}
  \caption{\label{fig1} Density of neutron states $n_{\kappa}(\varepsilon)$ for different blocks in $^{208}{\rm Pb}$ obtained by solving a Dirac equation with Woods-Saxon mean-filed potentials by Green's function method (blue solid line) , in comparison with the results for free neutrons calculated with potentials $V=S=0$ (red solid line). The black-dashed line at $\varepsilon=0$~MeV indicates the position of the continuum threshold.}
\end{figure*}

\section{Numerical details}
\label{sec:Numer}

In this work, taking the double-magic nucleus $^{208}$Pb as an example, the PSS and SS in the single-particle resonant states are investigated for neutrons.
The mean-field potentials in the Dirac equation~(\ref{EQ:RDirac}) are given in the Woods-Saxon form~\cite{PRL2001Alberto_86_5015,PRC2005Guo_72_054319,PRC2013BNLv_88_024323}
\begin{eqnarray}
  &&\Sigma(r)= \frac{\Sigma_0}{1+{\rm exp}[(r-R)/a]},
  \nonumber\\
  &&\Delta(r)= \frac{\Delta_0}{1+{\rm exp}[(r-R)/a]},
\end{eqnarray}
where $\Sigma_{0}=-66$~MeV, $\Delta_{0}=650$~MeV, $R=7$~fm, and $a=0.6$~fm.
As discussed in Ref.\cite{PRL2001Alberto_86_5015}, this potential is realistic enough to study the splittings of pseudospin partners, though it is not fully self-consistent.

The equations in the GF-RMF model are solved in the coordinate space with a mesh step of $0.1$~fm and a cutoff of $R_{\rm box}=20$~fm. To calculate the density of states $n_{\kappa}(\varepsilon)$, the parameter $\epsilon$ in Eq.~(\ref{EQ:DOS-lj}) is taken as $1\times10^{-6}$~MeV and the energy step $d\varepsilon$ is $1\times10^{-4}$~MeV. With this energy step, the accuracy for energies and widths of the single-particle resonant states can be up to $0.1$~keV.

\section{Results and discussion}
\label{sec:resu}

In Fig.~\ref{fig1}, the density of states $n_{\kappa}(\varepsilon)$ in different blocks $\kappa$ for neutrons in $^{208}$Pb are plotted as a function of the single-particle energy $\varepsilon$. The dotted line in each panel indicates the continuum threshold. The peaks of $\delta$-functional shapes below the continuum threshold correspond to bound states and spectra with $\varepsilon>0$ are continuous. In the continuum, by comparing the density of states for $^{208}$Pb (denoted by blue solid line) and those for free particles obtained with zero potentials $\Sigma(r)=\Delta(r)=0$ (denoted by the red solid line), one can easily find out the resonant states. It is clear that the density of states $n_{\kappa}(\varepsilon)$ for the resonant states sit atop of those for free particles. Accordingly, the neutron resonant states are observed in $d_{3/2}$, $d_{5/2}$, $g_{7/2}$, $h_{9/2}$, $h_{11/2}$, $i_{13/2}$, $j_{13/2}$, $j_{15/2}$ and $k_{15/2}$ blocks. Among these resonant states, the states $h_{9/2}$ and $i_{13/2}$ are relatively very wide.

\begin{figure}[!t]
  \includegraphics[width=0.9\linewidth]{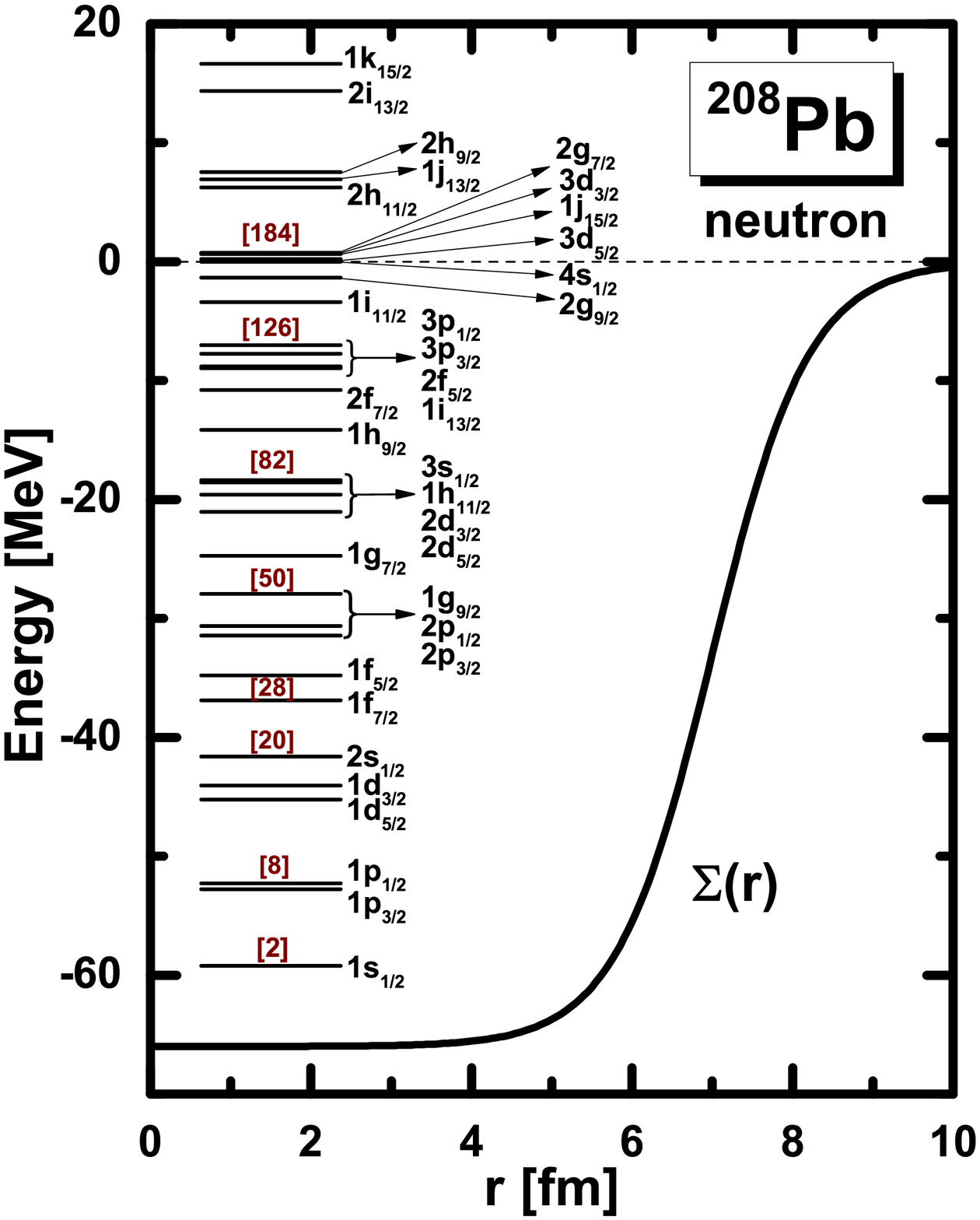}
  \caption{\label{fig2} Single-particle levels and the mean-field potential $\Sigma(r)$ for neutrons in $^{208}$Pb. The neutron magic numbers 2, 8, 20, 28, 50, 82, and 126 and the candidate 184 are given.}
\end{figure}

\begin{table}[htp!]
\center
\caption{Energies $\varepsilon_{\rm res}$ and widths $\Gamma$ of single neutron resonant states for $^{208}$Pb in the GF-RMF calculations. Data are in units of MeV.}
\label{Tab1}
\begin{tabular}{ccccccc}
  \hline\hline
  \multirow{2}*{$nl_{j}$ }& \multicolumn{2}{c}{positive parity } &~~~& \multirow{2}*{$nl_{j}$ } &  \multicolumn{2}{c}{negative parity }\\
           \cline{2-3}\cline{6-7}
            &$\varepsilon_{\rm res.}$&$\Gamma$ & &          &$\varepsilon_{\rm res.}$&$\Gamma$ \\\hline
  $3d_{3/2}$ &0.704&0.358 &&$2h_{9/2}$  &7.533&2.520\\
  $3d_{5/2}$ &0.221&0.108 &&$2h_{11/2}$ &6.261&1.630\\
  $2g_{7/2}$ &0.776&0.012 &&$1j_{13/2}$ &6.926&0.100\\
  $2i_{13/2}$&14.343&5.200&&$1j_{15/2}$ &0.648&0.010\\
  $1k_{15/2}$&16.629&1.978&&&&\\
 \hline\hline
\end{tabular}
\end{table}

From the density of states, the energies of the bound states as well as the energies ($\varepsilon_{\rm res.}$) and widths ($\Gamma$) of the resonant states can be extracted~\cite{PRC2014TTSun_90_054321}. In Table.~\ref{Tab1}, we list the energies and widths of the single-neutron resonant states in $^{208}$Pb. In Fig.~\ref{fig2}, the mean-field potential $\Sigma(r)$ and the corresponding single-particle levels including both the bound states and resonant states are shown for neutrons in nucleus $^{208}$Pb. A good shell structure of the single-particle spectra can be observed. The traditional neutron magic numbers 2, 8, 20, 28, 50, 82 and 126 and the candidate 184~\cite{NPA2005WZhang_753_106} are well reproduced.

\begin{figure}[htbp]
  \includegraphics[width=0.9\linewidth]{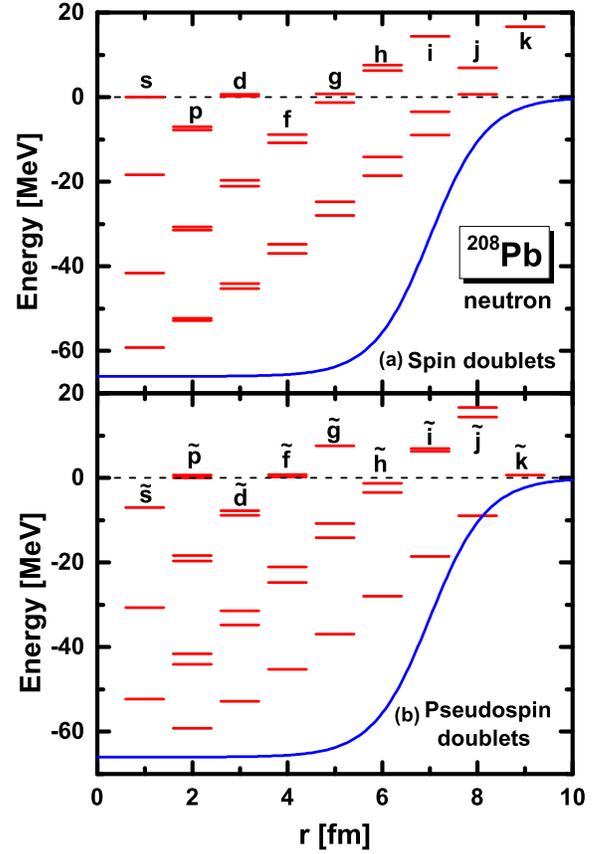}
  \caption{\label{fig3} Spin doublets (a) and pseudospin doublets (b) of the single-neutron spectra in $^{208}$Pb.}
\end{figure}

In Fig.~\ref{fig3}, we present the single-neutron spectra in nucleus $^{208}$Pb in forms of spin doublets and pseudospin doublets with respect to the angular momentum $l$ and pseudoangular momentum $\tilde{l}$, as well as the mean-field potential $\Sigma(r)$. Four spin doublets $3d$, $2g$, $2h$, and $1j$ and also four pseudospin doublets $3\tilde{p}$, $2\tilde{f}$, $1\tilde{i}$, and $1\tilde{j}$ are observed for the single-neutron resonant states. However, the state $2g_{9/2}$ of $2g$ spin doublet and the state $4s_{1/2}$ of $3\tilde{p}$ pseudospin doublet are weakly bound states. For the spin doublets in Fig.~\ref{fig3}(a), SO splittings between the doublets with low angular momenta $l$ are relatively small, such as the $p$ and $d$ orbits. Besides, the splittings of the doublets with the same main quantum number ($n=1$ or $n=2$) increase with the angular momentum $l$, except the $2h$ resonant doublets locating in the continuum.
This monotonic relation can be explained by the centrifugal barrier ${\displaystyle V_{\rm CB}=\frac{1}{M_+}\frac{l(l+1)}{r^2}}$ keeping the particle away from the center so that a big overlap between the wave functions and the spin-orbit potential always happens for larger $l$~\cite{NPA1999JMeng_650_176}.
For all the pseudospin doublets with $\tilde{l}>0$ in Fig.~\ref{fig3}(b), there is always one state without pseudospin partners. These states are simply the intruder states with $\kappa<0$, i.e., pseudospin antialigned. The appearance of these intruder state has been explained in a novel way with the Jost function~\cite{PRC2013BNLv_88_024323}, which is related to the lower components of Dirac wave functions. The number of zeros of Jost functions of pseudospin antialigned states is always one more than that of pseudospin aligned states, hence one intruder state remains. This intruder state has also been naturally emerged in supersymmetric quantum mechanics ~\cite{PRC2013Liang_87_014334,PRC2013Shen_88_024311,PRL2004Leviatan_92_202501,PRL2009Leviatan_103_042502}.
Meanwhile, different from the spin doublets, the pseudospin partners perform an obvious threshold effect. For example, the pseudospin partners $3\tilde{p}$ ($3d_{3/2}$ and $4s_{1/2}$), $2\tilde{f}$ ($2g_{7/2}$ and $3d_{5/2}$), and $1\tilde{i}$ ($1j_{13/2}$ and $2h_{11/2}$)~are very close to the threshold and their PSO splittings are as small as 0.5~MeV.

\begin{figure}[!t]
  \includegraphics[width=0.9\linewidth]{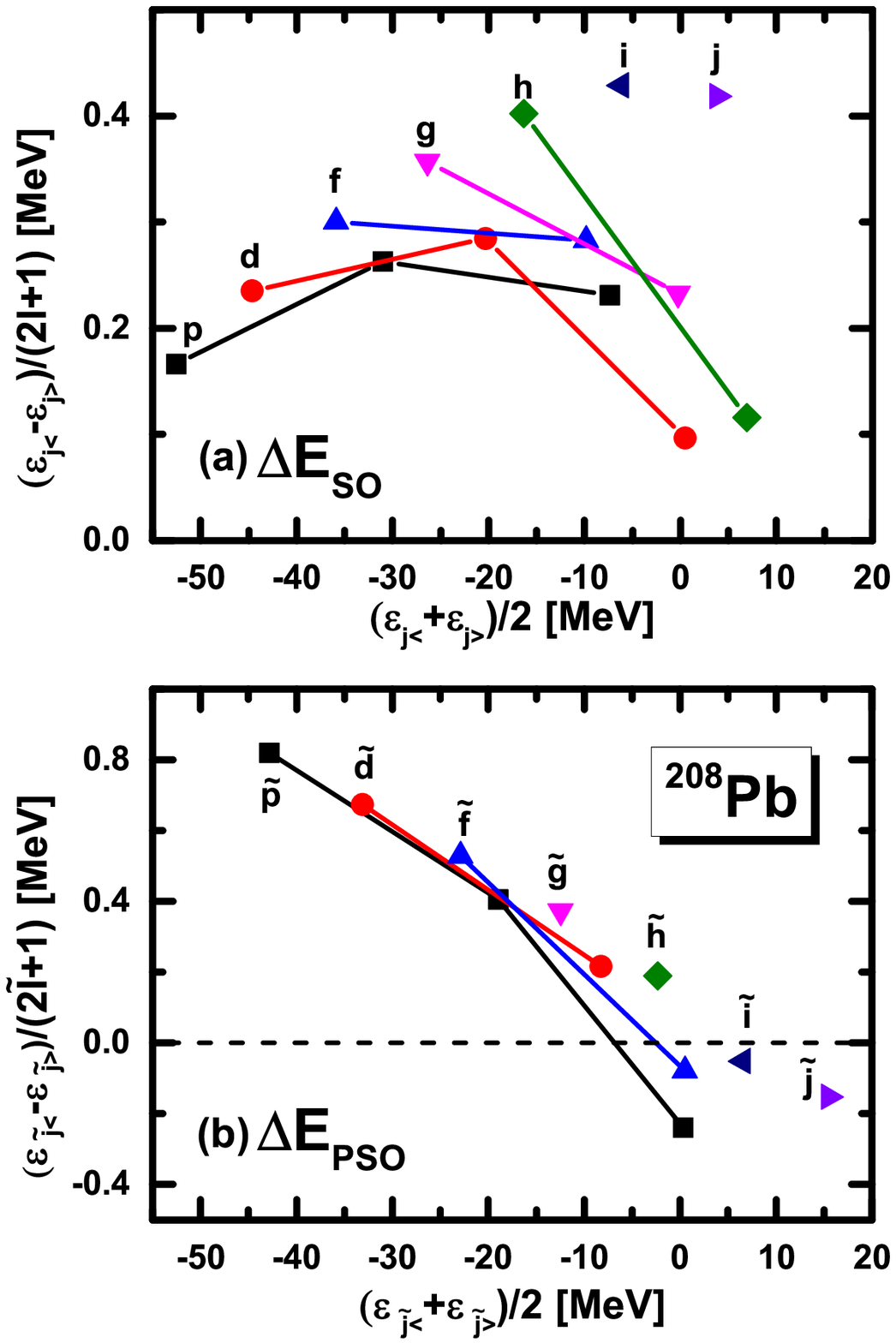}
  \caption{\label{fig4} Reduced SO splitting $\Delta E_{\rm SO}=(\varepsilon_{j_<}-\varepsilon_{j_>})/(2l+1)$ (a) and reduced PSO splitting $\Delta E_{\rm PSO}=(\varepsilon_{\tilde{j}_<}-\varepsilon_{\tilde{j}_>})/(2\tilde{l}+1)$ (b) versus their average single-particle energies $E_{\rm av}=(\varepsilon_{j_<(\tilde{j}_<)}+\varepsilon_{j_>(\tilde{j}_>)})/2$ in single-neutron spectra of $^{208}$Pb. For the spin doublets, $j_<=l-1/2$ and $j_{>}=l+1/2$, and for the pseudospin doublets, $\tilde{j}_<=\tilde{l}-1/2$ and $\tilde{j}_>=\tilde{l}+1/2$. Spin(pseudospin) doublets with the same $l$($\tilde{l}$) are linked by lines.}
\end{figure}

To show the SO and PSO splittings more clearly, the reduced SO splittings $\displaystyle \Delta E_{\mathrm{SO}} = (\varepsilon_{j_<}-\varepsilon_{j_>})/(2l+1)$ and reduced PSO splittings $\displaystyle \Delta E_{\mathrm{PSO}} = (\varepsilon_{\tilde{j}_<}-\varepsilon_{\tilde{j}_>})/(2\tilde{l}+1)$ versus the average single-particle energies $E_{\rm av}=(\varepsilon_{j_<(\tilde{j}_<)}+\varepsilon_{j_>(\tilde{j}_>)})/2$ are respectively plotted in Figs.~\ref{fig4}(a) and \ref{fig4}(b). From Fig.~\ref{fig4}(b), a dramatic energy dependence of PSS is observed. The energy splittings $\Delta E_{\mathrm{PSO}}$ of $3\tilde{p}$ ($3d_{3/2}$ and $4s_{1/2}$), $2\tilde{f}$ ($2g_{7/2}$ and $3d_{5/2}$), $1\tilde{i}$ ($1j_{13/2}$ and $2h_{11/2}$), and $1\tilde{j}$ ($1k_{15/2}$ and $2i_{13/2}$) even become negative, which means the orders of the doublets have reversed. This reversed level structure is decided by the sign of the integration of the pseudospin-orbit potential $\displaystyle (-\frac{1}{M_{-}^{2}}\frac{d M_{-}}{d r}|F|^{2})$ over $r$~\cite{PRC1999JMeng_59_154}. And if the pseudospin-orbit potential vanishes, the PSS will be exactly conserved. This can also be explained by the spin-orbit effects within the framework of supersymmetric quantum mechanics as discussed in Refs.~\cite{PRC2013Liang_87_014334,PRC2013Shen_88_024311}.

\begin{figure}[t!]
  \includegraphics[width=0.9\linewidth]{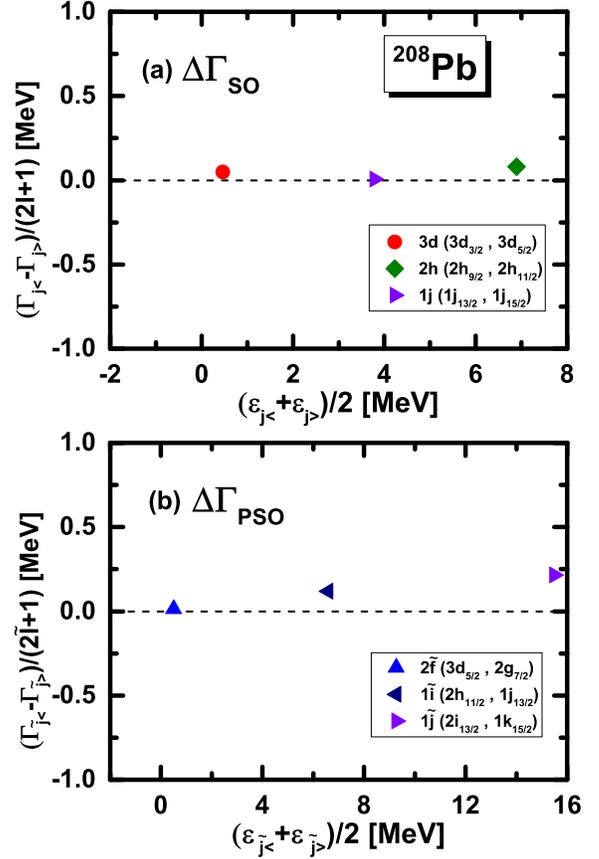}
  \caption{\label{fig5} The same as Fig.~\ref{fig4} but for the reduced SO width splitting $\Delta \Gamma_{\rm SO}=(\Gamma_{j_<}-\Gamma_{j_>})/(2l+1)$ (a) and the reduced PSO width splitting $\Delta \Gamma_{\rm PSO}=(\Gamma_{\tilde{j}_<}-\Gamma_{\tilde{j}_>})/(2\tilde{l}+1)$ (b).}
\end{figure}

For resonant states, the width is another significant feature. Thus we also study the width splittings for the spin doublets and pseudospin doublets for the resonant states.
In Fig.~\ref{fig5}, the reduced SO width splittings
$\displaystyle \Delta \Gamma_{\rm SO}=(\Gamma_{j_<}-\Gamma_{j_>})/(2l+1)$
and reduced PSO width splittings
$\displaystyle \Gamma_{\rm PSO} = (\Gamma_{\tilde{j}_<}-\Gamma_{\tilde{j}_>})/(2\tilde{l}+1)$
versus their average single-particle energies
$E_{\rm av}=(\varepsilon_{j_<(\tilde{j}_<)}+\varepsilon_{j_>(\tilde{j}_>)})/2$ are presented for the single-neutron resonant states in $^{208}$Pb.
Different from the case of the energy splittings, the width splittings $\Gamma_{\rm(P)SO}$ both for the spin and pseudospin doublets keep positive, which means that the width splittings of the doublets are not reversed.
For the spin doublets, because of the spin-orbit coupling, the single-particle energy of the state with positive $\kappa$ are larger than that of the state with negative $\kappa$, but their centrifugal-barriers are approximately identical. Therefore, the widths of the states with positive $\kappa$ are larger. For the pseudospin doublets, as pointed in Ref.~\cite{PRC2005JYGuo_72_054319}, such positive width splittings are mainly caused by the higher centrifugal barrier of the larger angular momentum state. Besides, the reduced SO width splittings are very small for all the three spin doublets locating from low energy to high energy, which are less than $0.1$~MeV. The same case happens for the low lying $1\tilde{f}$ pseudospin doublet. However, the width splitting for the pseudospin doublets $1\tilde{j}$ ($2i_{13/2}$ and $1k_{15/2}$) is relatively larger. This is mainly contributed by the rather large resonant width of $2i_{13/2}$ ($5.20$~MeV), while the small width of its partner $1k_{15/2}$ ($1.98$~MeV).

Comparing Fig.~\ref{fig4} and Fig.~\ref{fig5}, there is no specific relationship between energy splittings and width splittings for the single-particle resonant (pseudo)spin doublets. The energy splitting of a pair of doublets could be very large, while the width splitting of them could be very small and even versa. For example, for the spin doublets $1j$, $\Delta E_{\rm SO}= 0.419$~MeV, while $\Delta \Gamma_{\rm SO}= 0.006$~MeV, and for the pseudospin doublets $1\tilde{i}$, $\Delta E_{\rm PSO}= -0.051$~MeV, while $\Delta \Gamma_{\rm PSO}= 0.118$~MeV. Systematic studies have been carried out in Ref.~\cite{PRC2005JYGuo_72_054319} for the splittings on the dependence of the shape of the Woods-Saxon potential, and it has been found that the depth, radius, and diffuseness of the potential play an important role in both the width and energy splittings of the resonant states.

\begin{figure}[!t]
  \includegraphics[width=0.9\linewidth]{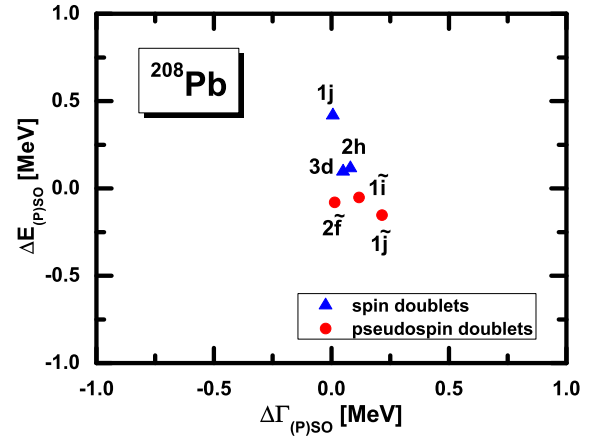}
  \caption{\label{fig6} Reduced energy splittings  $\Delta E_{\rm (P)SO}$ versus reduced width splittings $\Delta \Gamma_{\rm (P)SO}$ for the (pseudo)spin doublets in $^{208}{\rm Pb}$.}
\end{figure}

\begin{figure}[!t]
  \includegraphics[width=0.9\linewidth]{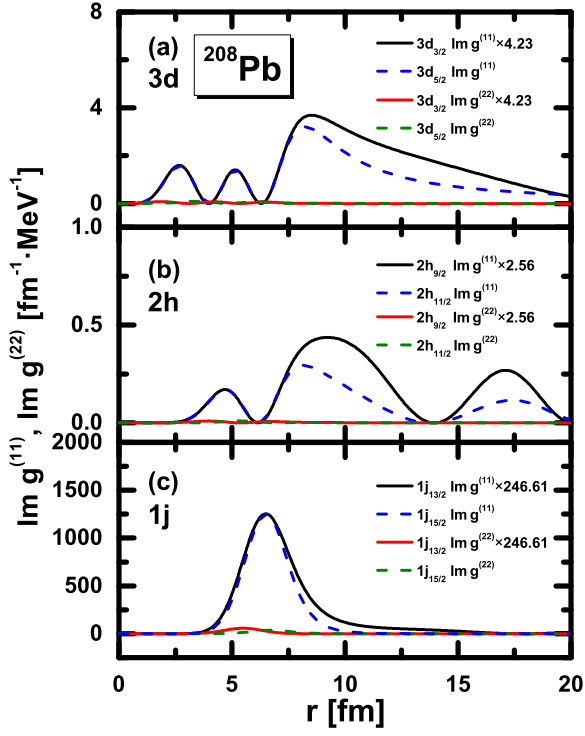}
  \caption{\label{fig7} Distribution function $\rho_g(r)={\rm Im}\mathcal{G}^{(11)}_{\kappa}(r,r;\varepsilon+i\epsilon)$ and $\rho_f(r)={\rm Im}\mathcal{G}^{(22)}_{\kappa}(r,r;\varepsilon+i\epsilon)$ in the coordinate space for the single-neutron resonance spin doublets (a)~$3d$, (b)~$2h$, and (c)~$1j$ in $^{208}{\rm Pb}$ at the resonant energy $\varepsilon=\varepsilon_{j_>(j_<)}$ extracted from $n(\varepsilon)$. For comparison, the functions for the spin doublets with smaller angular momenta $j_<$ are  multiplied by a factor of $4.23, 2.56$, and $246.61$, respectively. With those factors, the density of states $n(\varepsilon)$ integrated by $\rho_g(r)+\rho_f(r)$ in Eq.~(\ref{EQ:DOS-lj}) at $\varepsilon=\varepsilon_{j_>(j_<)}$ have the same hight.}
\end{figure}

In order to investigate better the (pseudo)spin symmetry in the single-particle resonant states, we present in Fig.~\ref{fig6} the reduced energy splittings $\Delta E_{\rm (P)SO}$ versus reduced width splittings $\Delta \Gamma_{\rm (P)SO}$ for the (pseudo)spin doublets in $^{208}{\rm Pb}$. It is seen that the $\Delta \Gamma_{\rm (P)SO}$ ranging from $0.0$~MeV to $0.25$~MeV are independent of $\Delta E_{\rm (P)SO}$.

\begin{figure}[t!]
  \includegraphics[width=0.9\linewidth]{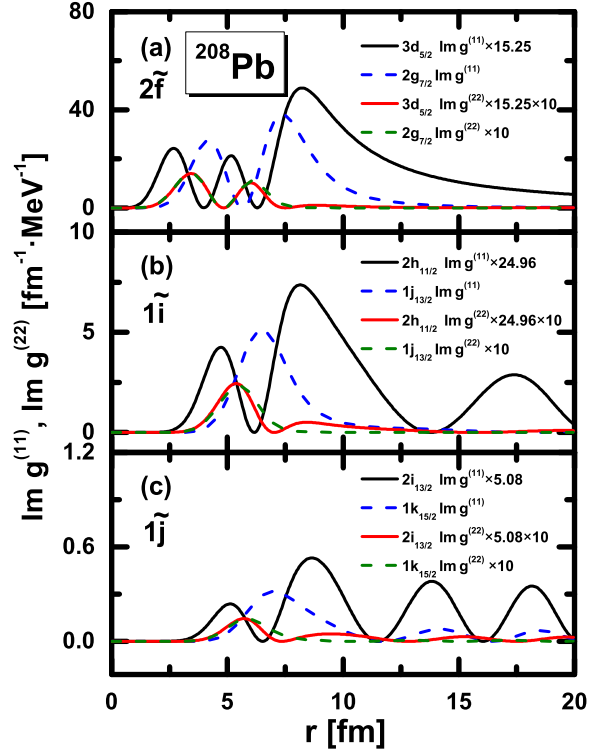}
  \caption{\label{fig8} The same as Fig.~\ref{fig7} but for the pseudospin doublets (a)~$3\tilde{p}$, (b)~$1\tilde{i}$, and (c)~$1\tilde{j}$ in $^{208}{\rm Pb}$. Besides, to depict the small components ${\rm Im}\mathcal{G}^{(22)}_{\kappa}(r,r;\varepsilon+i\epsilon)$ clearly, all of them are enlarged by 10 times.}
\end{figure}

The Dirac wave functions for the (pseudo)spin doublets provide another way to check the (pseudo)spin approximation in nuclei~\cite{PRC1998Ginocchio_57_1167}, i.e., for the spin doublets, the upper components of the Dirac spinor $G(r)$ are similar, and for the pseudospin doublets, the lower components of the Dirac spinor $F(r)$ are similar. In Ref.~\cite{CPL2007SSZhang_24_1199}, such similarity properties for resonant states have been discussed by ACCC method. In the framework of Green's function method, the wave functions for the (pseudo)spin doublets are not calculated directly. However, it can be inferred from Eqs.~(\ref{EQ:GF}) and (\ref{EQ:DOS-lj}) that the density distribution functions $\rho_g(r)={\rm Im}\mathcal{G}_{\kappa}^{(11)}(r,r;\varepsilon+i\epsilon)$ and $\rho_f(r)={\rm Im}\mathcal{G}_{\kappa}^{(22)}(r,r;\varepsilon+i\epsilon)$ have a certain relationship with the wave functions, which are the integrands for the density of states $n_{\kappa}(\varepsilon)$ and correspond to the upper Dirac spinor $G(r)$ and lower Dirac spinor $F(r)$, respectively. Therefore, we show in Figs.~\ref{fig7} and \ref{fig8} the density distribution functions $\rho_g(r)$ and $\rho_f(r)$ of different single-particle orbits.

Fig.~\ref{fig7} shows the density distributions of the spin doublets $3d$, $2h$, and $1j$ in $^{208}{\rm Pb}$. It can be seen that the integrands $\rho_g(r)={\rm Im}\mathcal{G}_{\kappa}^{(11)}(r,r;\varepsilon+i\epsilon)$ of each doublets are quite similar and almost identical in the region around $r<7.5$~fm, where the nuclear potential dominates. In Fig.~\ref{fig8}, we show the density distributions of the pseudospin doublets $3\tilde{p}$, $1\tilde{i}$, and $1\tilde{j}$. As seen in the figure, for the pseudospin doublets, the integrands $\rho_f(r)={\rm Im}\mathcal{G}_{\kappa}^{(22)}(r,r;\varepsilon+i\epsilon)$ of each doublets are also quite similar and almost identical within the nuclear potential region. In conclusion, we can deduce with GF method that the similarity properties of the wave functions of the spin and pseudospin doublets still maintain in resonant states.

\section{Summary}
\label{sec:Sum}

In this work, taking the double-magic nucleus $^{208}$Pb as an example, the SS and PSS in the single-neutron resonant states are studied with the Green's function method. The mean-field potential is taken in a Woods-Saxon form, which can well reproduce the traditional magic numbers and is realistic enough to study the splittings of pseudospin partners. Three spin doublets $3d$, $2h$, and $1j$ and also three pseudospin doublets $3\tilde{p}$, $1\tilde{i}$, and $1\tilde{j}$ are obtained for the single-neutron resonant states in $^{208}$Pb.

First, the energy splittings for the (pseudo)spin doublets are investigated. It is found that the SO splittings with the same main quantum number increase with the angular momentum. The pseudospin doublets perform an obvious threshold effect, i.e., the energy splittings of doublets locating around the zero energy are very small. Moreover, a dramatic energy dependence of PSS is observed and the orders of the doublets are reversed for the resonant states.

Second, the width splittings for the (pseudo)spin doublets are investigated. We find that the reduced width splittings of either the spin doublets or the pseudospin doublets are systematically larger than zero. 
The width splittings are very small ranging $0$~MeV to $0.25$~MeV, which are independent with the energy splittings.

Finally, as an examination, we check the similarity of the distribution functions $\rho_{g}(r)$ and $\rho_{f}(r)$ which respectively correspond to the upper component $G(r)$ and the lower component $F(r)$ of the Dirac spinor for the (pseudo)spin doublets. For the spin doublets, $\rho_{g}(r)$ are quite similar, while for the pseudospin doublets $\rho_{f}(r)$ are quite similar. We conclude that the similarity properties of the wave functions of the spin and pseudospin doublets still maintain in resonant states.

\begin{acknowledgements}
This work was supported by the National Natural Science Foundation of China (Grant No.~11505157).
\end{acknowledgements}

\end{document}